\documentclass[twoside,11pt]{article}

\usepackage{obs_study_style}
\usepackage{graphicx}
\usepackage{enumerate}
\usepackage{url} % not crucial - just used below for the URL 
\usepackage{import}
\usepackage{amssymb, amsmath, latexsym, bm, xcolor}
\usepackage{multirow}

\newtheorem{assumption}{Assumption}

\heading{}{}{}{For Review}{Not Yet}{Yumin Zhang and Arman Sabbaghi}

\ShortHeadings{Distributed Design for Causal Inferences on Big Observational Data}{Zhang and Sabbaghi}
\firstpageno{1}

\begin{document}

\title{Distributed Design for Causal Inferences on Big Observational Data}

\author{\name Yumin Zhang \email zhan2013@purdue.edu \\
       \addr Department of Statistics\\
       Purdue University\\
       West Lafayette, IN 47907, USA 
       \AND
       \name Arman Sabbaghi \email sabbaghi@purdue.edu \\
       \addr Department of Statistics\\
       Purdue University\\
       West Lafayette, IN 47907, USA}

\maketitle

\begin{abstract}%   <- trailing '%' for backward compatibility of .sty file
A fundamental issue in causal inference for Big Observational Data is confounding due to covariate imbalances between treatment groups. This can be addressed by designing the data prior to analysis. Existing design methods, developed for traditional observational studies with single designers, can yield unsatisfactory designs with suboptimum covariate balance for Big Observational Data due to their inability to accommodate the massive dimensionality, heterogeneity, and volume of the Big Data. We propose a new framework for the distributed design of Big Observational Data amongst collaborative designers. Our framework first assigns subsets of the high-dimensional and heterogeneous covariates to multiple designers. The designers then summarize their covariates into lower-dimensional quantities, share their summaries with the others, and design the study in parallel based on their assigned covariates and the summaries they receive. The final design is selected by comparing balance measures for all covariates across the candidates. We perform simulation studies and analyze datasets from the 2016 Atlantic Causal Inference Conference Data Challenge to demonstrate the flexibility and power of our framework for constructing designs with good covariate balance from Big Observational Data.
\end{abstract}

\begin{keywords}
  Matching; Propensity score; Subclassification
\end{keywords}

\section{Introduction}
\label{sec-intro}

Observational studies are critical for causal inference. They are typically more prevalent, easier to construct, and broader in scope than randomized experiments. In contrast to designed experiments, the mechanism by which treatments are assigned to experimental units or subjects is not specified or known in an observational study. This distinguishing feature of an observational study typically leads to imbalances in covariates between their different treatment groups, which yields biases in causal inferences on treatment effects \citep{R2008}. The principle of designing an observational study before the analysis so as to approximate a randomized experiment can reduce biases in their causal inferences \citep{R2002, Rubin2006, IR2015}. This is because the explicit process of designing an observational study so as to yield treatment groups with similar covariate distributions directly reduces severe covariate confounding and generally leads to more valid causal conclusions than those obtained from analyses of the raw observational data \citep{S2010}. Common design methodologies for traditional observational studies include matching and subclassification \citep{RR1983, Rubin2006, S2010, IR2015}. Observational studies can ultimately be effective for obtaining valid causal inferences when they are first designed prior to their analyses. 

We refer to large observational data in the modern data science era as \textit{Big Observational Data}. Such data are increasing in prevalence due to advances in technologies (e.g., Internet devices and sensors) that make it possible to measure a wider variety of information on subjects compared to traditional observational studies. Besides its large volume, a Big Observational Dataset could typically constitute high-dimensional covariates where information of heterogeneous characteristics is encoded. For example, Purdue University built the Institutional Data Analytics Platform (IDAP) to store a tremendous amount of student data collected both from traditional demographics, course registration, and grade databases as well as from cutting-edge measurements on student wireless network activities, campus locations across the day, and access to specified campus buildings \citep{IDAP}. An objective for the IDAP is to infer the causal effects of student behaviors and activities on academic success. The IDAP constitutes Big Observational Data because it consists of high-dimensional covariates on heterogeneous types of information, a large volume of subjects, and the mechanism underlying the assignment of different interventions to the subjects (i.e., student behaviors and activities) was not specified or known by design. 

Existing design methodologies for achieving good covariate balance from traditional observational studies are generally ineffective for Big Observational Data. The high-dimensional covariates make it difficult for an individual analyst to design the data and assess the balance of each covariate across treatment groups. For example, traditional propensity score-based methods such as those that follow the paradigm of \citet{RR1983} cannot accommodate the high-dimensionality of Big Observational Data. \citet{DDFLS2020} recently demonstrated how such methods can fail due to lack of overlap for high-dimensional covariates across treatment groups. Big Observational Data also introduce challenges in computation and data privacy when there is only a single analyst. This is especially true for the data in the IDAP, which contains a large amount of sensitive information on more than a hundred thousand undergraduates over multiple years at Purdue University. One interesting approach to address these failures of existing methods is the proposal of \cite{LVKF2016} to randomly project covariates into a lower dimensional space and conduct nearest neighbor matching for the projection. A limitation of this approach is that it does not involve explicit assessments of covariate balance, which generally impairs the ability of the procedure to reduce biases in causal inferences. Recently, \citet{BVZ2020} tackles the problem of matching by directly optimizing the covariate balance via mixed-interger programming (MIP). They developed a linear-sized formulation of the MIP problem to address the heavy computation due to large observational datasets. However, the formulation may not work under high-dimensional covariates. Furthermore, the heterogeneous covariates in Big Observational Data can require different types of domain expertise in the design and analysis of the study. Thus, Big Observational Data are distinguished from traditional observational studies in that their high-dimensional and heterogeneous covariates, as well as large volume, make it infeasible for one individual to use existing design methodologies to obtain a designed observational study with decreased covariate imbalance, and thereby obtain valid causal inferences from them.

We propose a distributed design framework for Big Observational Data. Our framework divides the covariates into separate sets, and assigns the sets to separate ``designers''. The designers can be actual people or computer algorithms. They take as their input the matrix of their assigned covariates and the treatments that the subjects received. The designers form lower-dimensional summaries of their assigned covariates and share their summaries with all other designers. Each designer then independently designs the study based on their assigned covariates and received summaries, and returns their designed dataset that satisfies a specified criterion for covariate balance. The final design that will be chosen for the analysis of the study is selected by comparing the multiple candidate designs in terms of covariate balance measures across the treatment groups. The division of covariates under our framework helps to reduce the heavy workload of designing the Big Observational Data, and directly addresses the challenges of high-dimensional and heterogeneous covariates, as well as large volume data. For example, instead of having a single designer inspect the balance of all the high-dimensional covariates, under our framework each designer would instead focus on their relatively lower-dimensional assigned covariates and summaries of the covariates shared by the other designers. Such a division of the work for designing the study also makes it possible for heterogeneous covariates to be appropriately processed, and for sufficient data privacy to be maintained.

We proceed in Section \ref{sec-bg}  to review the notations and assumptions for our framework. Section \ref{sec-method} contains the description of our distributed design framework for Big Observational Data. We illustrate the frequentist properties for covariate balance under our design framework via simulation studies in Section \ref{sec-simu}. We provide further demonstrations of the utility of our framework for achieving covariate balance in Big Observational Data using thousands of datasets from the 2016 Atlantic Causal Inference Conference (ACIC) Data Challenge \citep{DHSSC2019} in Section \ref{sec-data}. Our concluding remarks are in Section \ref{sec-conclusion}.

\section{Background}
\label{sec-bg}

\subsection{Notation}
\label{sec:notations}

In this article we consider observational studies with two levels of a single treatment factor under the Rubin Causal Model \citep[RCM,][]{Holland1986}. For any observational study, we let $N$ denote the number of experimental units or subjects, $X_i \in \mathbb{R}^p$ denote the vector of observed covariate for subject $i$, and $W_i \in \{0, 1\}$ denote the treatment indicator for subject $i$. Each covariate vector includes all covariates measured or observed prior to treatment assignment, but not their transformations or interactions. Level $0$ of $W_i$ indicates the control and level $1$ indicates the active treatment. For the Science of our observational studies \citep{R2005} we assume that subjects comply or adhere to their treatment assignment, there are no principal strata \citep{FrangakisRubin2004}, and that the Stable Unit-Treatment Value Assumption \citep[SUTVA,][p.~9--13]{IR2015} holds. Under SUTVA, there are no lurking varieties of treatments that give rise to different potential outcomes, and there is no interference amongst subjects. Under these assumptions we denote the potential outcomes for subject $i$ under control and treatment by $Y_{i}(0)$ and $Y_{i}(1)$, respectively. The observed outcome $Y_i^{\mathrm{obs}}$ for subject $i$ is a function of $W_i$ and their potential outcomes, namely, $Y_i^{\mathrm{obs}} = W_iY_i(1) + (1-W_i)Y_i(0)$. Causal effects under the RCM are defined as comparisons of potential outcomes for a set of subjects \citep[p.~5--7]{IR2015}. Specifically, the standard individual treatment effect $\tau_{i}$ for subject $i$ is $\tau_{i} = Y_{i}(1) - Y_{i}(0)$ and the finite-population average treatment effect $\tau_{\mathrm{ATE}}$ is the average of all the individual treatment effects, i.e., $\tau_{\mathrm{ATE}}= \sum_{i=1}^{N}\tau_i/N$.

We denote the number of designers in our distributed design framework by $M$. Data and quantities associated with designer $m$ are indicated by superscripts. For example, $X_{i}^{(m)}$ indicates the vector of the $p^{(m)}$ covariates for subject $i$ that were assigned to designer $m$ ($1 \leq p^{(m)} < p$ for all $m = 1, \ldots, M$), and $X^{(m)}$ indicates the $N \times p^{(m)}$ matrix of covariates on all the subjects that are made available to designer $m$. 

\subsection{Assumptions for the Treatment Assignment Mechanism}
\label{sec:assumptions_treatment}

The treatment assignment mechanism is the process by which subjects receive treatment or control in a study \citep[p.~31]{IR2015}. More formally, it is the probability distribution for the vector of treatment indicators for all the subjects conditional on the covariates and potential outcomes. Knowledge of the assignment mechanism is critical to obtaining valid causal inferences, whether for randomized experiments or observational studies. \citet[p.~31--44]{IR2015} provide detailed explanations of treatment assignment mechanisms under the RCM. For the observational data that we consider, we will assume their treatment assignment mechanisms satisfy the following three properties. Further details on these properties are provided by \citet[p.~37--39]{IR2015}. In these assumptions $X$ denotes the $N \times p$ matrix of covariates for all subjects, $W \in \{0, 1\}^N$ denote the vector of treatment indicators, and $Y(0), Y(1) \in \mathbb{R}^N$ denote the vectors of control and treatment potential outcomes, respectively, for the subjects.

\begin{assumption}[Probabilistic]
\label{assump-probabilistic}
For each subject $i$, $0 < \mathrm{Pr} \{ W_{i} = 1 \mid  X, Y(0), Y(1) \} < 1$. 
\end{assumption}

\begin{assumption}[Unconfounded]
\label{assump-unconfounded}
For any $w \in \{0, 1\}^N$ and potential outcome vectors $Y(0), Y'(0), Y(1), Y'(1) \in \mathbb{R}^N$, $\mathrm{Pr} \{ W = w \mid  X, Y(0), Y(1) \} = \mathrm{Pr} \{ W = w \mid  X, Y'(0), Y'(1) \}$. 
\end{assumption}

\begin{assumption}[Individualistic]
\label{assump-individualistic}
A function $q: \mathbb{R}^{p+2} \rightarrow (0,1)$ exists so that for all subjects $i$, $\mathrm{Pr} \{ W_i = 1 \mid  X, Y(0), Y(1) \} = q(X_i, Y_i(0), Y_i(1))$ and $\mathrm{Pr} \{ W_i = 1 \mid  X, Y(0), Y(1) \} = c \prod_{i=1}^N q(X_i, Y_i(0), Y_i(1))^{W_i}\{1-q(X_i,Y_i(0),Y_i(1))\}^{1-W_i}$, where $c$ is the normalization constant for the probability mass function of the assignment mechanism.
\end{assumption}
The probabilistic assumption is also referred to as the overlap assumption. It permits the consideration of all subjects in the data as candidates for the design and analysis of the observational study. It also reduce the risk of extrapolation when estimating treatment effects for subjects with extreme probabilities \citep[p.~262]{IR2015}. Assumption \ref{assump-unconfounded} implies that there are no lurking confounders in the observational study, i.e., the observed covariates contain all the information governing treatment assignment and no additional variables associated with the outcomes are related to treatment assignment. If the individualistic assumption is not satisfied, then some subjects' treatment assignments would be dependent on the covariates and potential outcomes of other subjects. This dependency would complicate the design of the observational study.

The combination of all three assumptions justifies designing the observational study so as to compare treated and control subjects with the same covariates and perform causal inferences. This is because if a treated and control subject have the same covariates, then their treatment assignments can be seen to have been performed in a random manner. Hence, comparing treated and control subjects with the same covariates should yield unbiased inferences for the treatment effects in the designed observational study. Under these three assumptions, the probability of treatment assignment for a subject is also referred to as its propensity score \citep[p.~35,~39]{RR1983, IR2015}, and denoted by the function $e: \mathbb{R}^p \rightarrow (0,1)$.

\subsection{Designing an Observational Study via Propensity Scores}
\label{background-matching}

The propensity score serves as an effective tool for designing traditional observational studies. This is because it can summarize covariates into a single number with the property that subjects with similar propensity scores would have similar covariate vectors on average \citep{RR1983}. This fact has led to the development of design methodologies for observational studies in which treated and control subjects with similar estimated propensity scores are either grouped together into subclasses~\citep{RR1984} or matched into pairs~\citep{RR1985}. Theory-wise, Rubin and Thomas~\citeyear{RT1992a, RT1992b, RT1996} provided properties for the propensity score matching, in the broader class of affinely invariant matching methods, when the covariate distributions are ellipsoidally symmetric. Generally, matching can be algorithmically implemented~\citep{R1989, R1991, Z2012, YSR2020} where the propensity score possibly can be used as a covariate, distance measure, or caliper. The vast literature on propensity score-based design methods can be found in the books of \citet{R2002, R2020}, \citet{Rubin2006}, and \citet{IR2015}, as well as in \citet{S2010}. Examples of its successful applications in yielding unbiased causal inferences can be found in the case studies of \citet{DW1999, R2001, I2005} and \citet{ER2015}. 

The particular propensity score-based design methods that we will consider in our simulation studies and demonstrations include both one-to-one matching and subclassification. The matching methods include nearest neighbor matching, calipar matching, and the technique recently developed by \citet{YSR2020} to conduct optimal matching for datasets with large numbers of experimental units. The method by \citet{YSR2020} reduces the amount of computation for optimal matching by imposing an optimally chosen caliper. As to the subclassification method, we implement the iterative procedure developed by \citet[p.~290--294]{IR2015}. The procedure divides data into two subclasses unless the treatment and control subjects are similar with respect to their propensity scores. Details of these design methods can be found in Appendix~\ref{appen:design-method}.

\subsection{Assessing Covariate Balance in Designed Studies}
\label{background-assessment}

Prior to analyzing a designed observational study, it is critical to confirm that the treated and control subjects have similar covariate distributions. This similarity is also known as covariate balance \citep[p.~309]{IR2015}. Various measures have been used to formally quantify covariate balance. Examples include the standardized difference in univariate moments and the Kolmogorov-Smirnov statistic for the control and treated subjects~\citep{A2009, Z2012}. It can be helpful to aggregate the balance measures of multiple covariates into a univariate summary to choose among candidate designs. Besides such quantitative covariate balance measures, domain knowledge can also be useful for designing specific studies, especially if there exist several candidate designs that achieve different types of covariate balance. For the observational studies that we consider, we focus on the maximum absolute standardized difference in means across all the covariates.

\section{Distributed Design Framework}
\label{sec-method}

\subsection{Outline of Framework}
\label{sec:framework_outline}

\begin{figure}
\centering
\includegraphics[width=\textwidth]{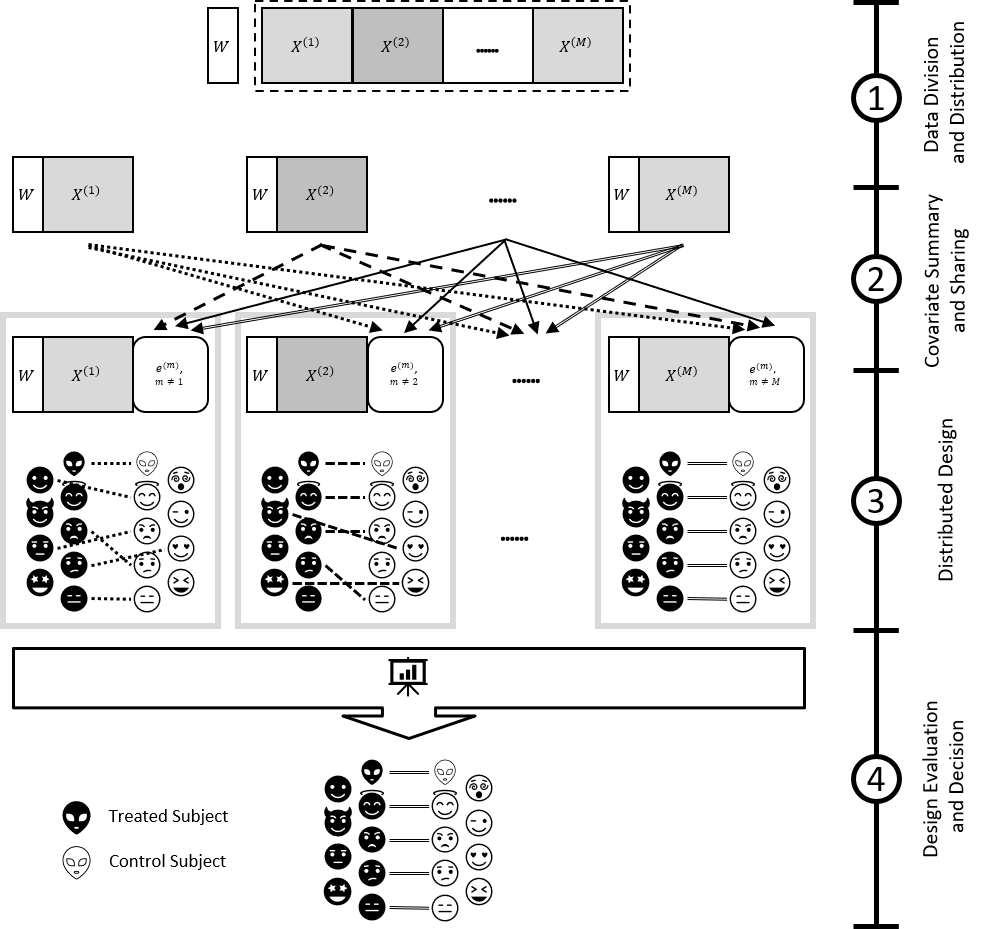}
\caption{Illustration of the steps involved in the distributed design framework for Big Observational Data. Different face icons represent subjects with potentially different characteristics while the colors denote the treatment groups.}
\label{fig:illustrative_figure}
\end{figure}

Our distributed design framework for Big Observational Data involves four general steps. First, the subjects' covariates are divided amongst the designers, and each designer is also given the treatment assignments for all subjects. Second, each designer calculates a summary of their assigned covariates for each subject, and shares their covariate summaries with the other designers. In this paper we consider one-dimensional summaries that are functions just of the assigned covariates. Third, each designer will design the observational study based just on their assigned covariates and the summaries they received from the other designers. Fourth, all candidate designs will be evaluated by means of a single, pre-specified covariate balance measure, and a single design will be chosen for the data analyses. In all these steps, no designer is given access to the subjects' observed outcomes. The outcomes will only be analyzed after the final design is chosen. Figure \ref{fig:illustrative_figure} illustrates these steps. Further details on the individual steps are provided in this section.

\subsection{Step 1: Data Division and Distribution}
\label{sec:framework_step_1}

In the first step the $N \times p$ matrix of covariates $X$ is divided according to its columns to form the designer-specific covariate matrices $X^{(1)}, \ldots, X^{(M)}$, with each $X^{(m)}$ having $p^{(m)}$ covariates selected from the $p$ total covariates. Each designer $m$ is given $X^{(m)}$ along with $(W_1, \ldots, W_N)^{\mathsf{T}}$. In practice, the $p^{(m)}$ covariates should be selected so that designer $m$ can focus on a smaller number of covariates from the higher-dimensional covariate space, and more carefully check the balance for their assigned covariates. Having $p^{(m)} \leq 20$ for each designer $m$ is typically feasible for designing a study.

The division and distribution of covariates can be performed in a flexible manner based on the characteristics of the covariates, the different designers' domain knowledge and expertise, the importance of certain covariates, and knowledge of how the covariates are related or interact with one another. For example, in the case of Purdue University's IDAP, students' demographic data and semester-specific course records can be assigned to different designers so that static and temporal covariates are appropriately and effectively considered in the design of the study. For covariate interactions that are significant for either the underlying treatment assignment mechanism or the potential outcomes model, the constituent covariates could be assigned as pairs to the designers.

\subsection{Step 2: Summarizing and Sharing Covariate Information}
\label{sec:framework_step_2}
A summary of the covariates that a designer $m$ can calculate is the estimate of a subject $i$'s propensity score based solely on the assigned covariates $X_i^{(m)}$. We denote the propensity score function for designer $m$ by $e^{(m)}(X_i^{(m)}) = \mathrm{Pr}(W_i = 1 \mid X_i^{(m)})$, and refer to it as the \emph{conditional propensity score} based only on the covariates assigned to designer $m$. The name ``conditional propensity score'' is based on the interpretation of $e^{(m)}(X_i^{(m)})$ as the propensity score conditioned on a subset of the covariates defined by the $X_i^{(m)}$. By virtue of the Tower Property of conditional expectation, the conditional propensity score is the projection of the standard propensity score onto the covariates in $X_i^{(m)}$, i.e., $e^{(m)}(X_i^{(m)}) = E\{e(X_i) \mid X_i^{(m)}\}$ where $e(X_i) = E(W_i \mid X_i^{(m)}, X_i^{(j)}, j \ne m)$ is the standard propensity score.

A designer is given great freedom in estimating their conditional propensity scores. They may freely choose from standard logistic regression-based approaches~\citep{RR1983}, machine learning approaches such as those based on boosting~\citep{MRM2004}, and other methods that are more suitable for the data under consideration. In any case, designer $m$ will share their estimated conditional propensity scores with the other designers. It is in this distributed manner that a designer obtains information from all covariates, while having to only process a smaller amount of covariate data themselves.

The concept of a conditional propensity score generally differs from that of the propensity score as provided by \citet{RR1983} because for the former a designer only considers their assigned covariates, instead of all observed covariates as for the latter. In certain cases estimates of propensity scores based on all the covariates can be recovered by a designer who utilizes their assigned covariates as well as the estimates of the conditional propensity scores shared with them by the other designers. To demonstrate one such case, assume that for each $m = 1, \ldots, M$ the columns of the covariate matrix $X^{(m)}$ are standardized to have mean $0$, and that for any two designers $m, m'$ their covariate matrices $X^{(m)}, X^{(m')}$ are orthogonal, i.e., $(X^{(m)})^{\mathsf{T}}X^{(m')}$ is the zero matrix of dimension $p^{(m)} \times p^{(m')}$. The first assumption is fairly standard in that designers are free to linearly transform their covariates prior to designing the study. The second assumption implies that the separate covariate matrices share minimum redundant information, and can be achieved in practice via careful assignment of the covariates to the designers. We further suppose that the conditional propensity scores for designer $m$ are modeled via ordinary least squares regression of $W$ onto the covariates in $X^{(m)}$. In this case, for designer $m$ we define the model
\begin{equation}
\label{eq:sliced_propensity_score_model}
W_i = \beta_0^{(m)} + \left ( X_i^{(m)} \right )^{\mathsf{T}}\beta^{(m)} + \epsilon_i,
\end{equation}
where $\epsilon_i$ is the Normal error term for standard linear regression. After fitting the model in equation (\ref{eq:sliced_propensity_score_model}) to the data, the conditional propensity score for subject $i$ is estimated by $\widehat{e}^{(m)}_i = \widehat{\beta}^{(m)}_0 + (X_i^{(m)})^{\mathsf{T}}\widehat{\beta}^{(m)}$. By standard regression calculations, the estimators of the coefficients in equation (\ref{eq:sliced_propensity_score_model}) are $\widehat{\beta}^{(m)}_0 = \sum_{i=1}^N W_i/N$ and $\widehat{\beta}^{(m)} = \{ (X^{(m)})^{\mathsf{T}}X^{(m)} \}^{-1}(X^{(m)})^{\mathsf{T}}W$ in this case. After the estimates of the conditional propensity scores are shared amongst the designers, we finally suppose that each designer estimates a final propensity score via ordinary least squares regression of $W$ onto the covariates in $X^{(m)}$ as well as the shared conditional propensity scores estimates. This final regression model is specified as
\begin{equation}
\label{eq:final_propensity_score_model}
W_i = \tilde{\beta}_0^{(m)} + \left ( X_i^{(m)} \right )^{\mathsf{T}}\tilde{\beta}^{(m)} + \sum_{j \neq m}\gamma_j^{(m)}\widehat{e}^{(j)}_i + \delta_i,
\end{equation}
where $\delta_i$ is the Normal error term. It follows by straightforward regression calculations that the estimated propensity scores obtained by fitting model (\ref{eq:final_propensity_score_model}) are equivalent to the estimated propensity scores obtained by fitting the model $W_i = \alpha_0 + X_i^{\mathsf{T}}\alpha + \xi_i$ to all of the covariate data, where $\xi_i$ is the Normal error term.

\subsection{Step 3: Distributed Design}
\label{sec:framework_step_3}

In the third step each designer utilizes the knowledge they have on their assigned covariates and the estimates of the conditional propensity scores obtained from the other designers to match or subclassify the subjects. These $M$ candidate designs for the observational study are constructed in parallel. Similar to the consideration in Step $2$, the specific design approaches can be determined according to each designers' data at hand. In our simulation studies and demonstrations we will have each designer $m$ estimate a new propensity score for each subject based on $X^{(m)}$ and their given estimated conditional propensity scores as predictors in the propensity score model. Each designer will then use their new propensity score estimates to create their designed study. Candidate design $m$ will be encoded by a vector $S^{(m)} \in \mathbb{Z}^N$, with integers in the vector indicating which subjects are put into specific matched pairs or subclasses, and which subjects are removed from consideration. For example, if we consider a study originally consisting of three treated and four control subjects with $W_1 = W_3 = W_5 = 1$ and $W_2 = W_4 = W_6 = W_7 = 0$, and for which designer $m$'s designed study has treated and control subjects placed into matched pairs and one control subject removed from consideration, then one encoding of this design is $S^{(m)} = (1,1,2,2,3,3,0)$. In this encoding,  $1, 2, 3$ denote the three matched pairs and $0$ indicates the subject removed from consideration. 

\subsection{Step 4: Design Evaluation and Decision}
\label{sec:framework_step_4}

The candidate designs $S^{(1)}, \ldots, S^{(M)}$ are finally shared with all designers to evaluate the balance for all covariates across all candidate designs in a distributed manner. Specifically, given a design proposal $S^{(m)}$, designer $m'$ will evaluate a pre-specified covariate balance measure for the $p^{(m')}$ covariates in $X^{(m')}$ and any of their important transformations or interactions, for all $m, m' = 1, \ldots, M$. In this manner each designer will evaluate just their assigned covariates. The evaluated covariate balance measures for all the candidates are pooled to identify the design that achieves optimum balance across all covariates.

\subsection{Flexibility of Framework}
\label{sec:framework_flexibility}

The general steps in our distributed design framework can accommodate a great deal of flexibility. For example, a select few important covariates can be shared in addition to conditional propensity score estimates for direct consideration by other designers in Step 2. The designers can also share constraints for the design of the observational study that others can incorporate when constructing their candidate designs. 

The amount of information communicated under this framework can be calculated in a straightforward manner. Sharing conditional propensity scores in Step $2$ and design candidates in Step $3$ incurs a transfer cost of $O(NM^{2})$, and sharing the covariate balance assessments in Step $4$ incurs a transfer cost of $O(pM)$. When the number of designers is fixed, the required information transfer efforts are proportional to the data volume, and our framework does not impose a sizeable burden to communication.

\section{Simulation Studies}
\label{sec-simu}

\subsection{Description of Simulation Settings}
\label{sec:simulation_description}

Each of our simulated datasets will have $N = 10000$ subjects and $p = 120$ covariates that are independent and identically distributed standard Normal random variables. Our choice of covariate distributions having mean $0$ and variance $1$ corresponds to the practice in randomized experiments and observational studies of standardizing covariates prior to data analyses. We consider the case of $M = 6$ designers throughout, with $p^{(m)} = 20$ for each $m = 1, \ldots, 6$ and $X^{(1)}, \ldots, X^{(6)}$ containing no covariates in common. 

The treatment assignment mechanism in our simulations will correspond to Bernoulli trials \citep[p.~47--50]{IR2015}, with $W_i \sim \text{Bernoulli}(p_{i})$ independently and with $\mathrm{log}\{p_i/(1-p_i)\}$ being a function of the covariates. Our specification of the $\mathrm{log}\{p_i/(1-p_i)\}$ for each dataset is done according to five steps that are meant to induce the types of complex covariate imbalances observed in real-life observational studies. The treatment assignment model includes polynomially-transformed terms with varying coefficients and interaction effect between the covariates. The details are provided in Appendix~\ref{appen:simu-data}.

We consider two different settings in which covariates are assigned to the designers. In the first setting, both parent covariates of an active two-factor covariate interaction will be assigned as a pair to a single designer. In the second setting, the parent covariates of an active two-factor covariate interaction can be assigned to distinct designers. We will see in the simulation study that important interactions generated under the second setting may not be accounted for in an effective manner in the design of the observational study. It is important to note that when a single designer identifies significant interactions and includes them in their conditional propensity score model, the other designers would be able to indirectly incorporate the interaction effects when they utilize the shared estimates of the conditional propensity scores to construct their candidate designs. 

The designers in our study will implement several propensity-score-based design methods. The propensity score is estimated via lasso logistic regression implemented in the glmnet package in R~\citep{SFHT2011}. Each designer will consider all their assigned covariates and given summaries, as well as the squared terms and linear-by-linear two-way interactions when fitting the model. The design methods include subclassification, nearest neighbor matching, caliper matching, and the optimal matching method proposed by~\cite{YSR2020}. Applying different design methods will provide better understanding of how our distributed design framework performs and show the flexibility of this design framework. Details of the design methods can be found in Appendix~\ref{appen:design-method}.

The designers in our study will estimate propensity scores via logistic regression and stratify subjects according to the \citet[p.~290--294]{IR2015} iterative stratification procedure. They will consider all their assigned covariates and given summaries as well as the squared terms and linear-linear two-way interactions using the lasso implemented in the glmnet package in R~\citep{SFHT2011}. In our implementation, the designers will discard extreme treated and control subjects whose estimated propensity scores lie beyond the range of those for the subjects in the other respective group. Also, they will split a stratum if the $p$-value of the two-sample $t$-test comparing the treated and control groups' mean propensity scores is less than $0.15$, the new strata have at least $50$ subjects, and either treatment group in the new strata have at least $30$ subjects. 

We consider two metrics to summarize all the covariates' balances for a candidate design. The first is the maximum absolute standardized difference in means between treatment groups: $d_{\text{max}}^{(m)} = \max_{j = 1, \ldots, p} \left\{d_{j}\left(S^{(m)}\right)\right\}$, where $d_{j}(S^{(m)})$ is the absolute standardized difference in means for covariate $j$ in candidate design $S^{(m)}$. The second is the number of covariates having their absolute standardized differences above the threshold of $0.2$: $d_{+}^{(m)} = \sum_{j = 1, \ldots, p} \mathbb{I}\left\{d_{j}\left(S^{(m)}\right)  > 0.2\right\}$. Our threshold of $0.2$ is based on the considerations in \citep[p.~340]{LLS2010}. There is no consensus on the threshold of the absolute standardized differences in means, as the choice depends on the specific context of an observational dataset~\citep{A2009, S2010, HSA2010}. We evaluate the candidate designs with respect to the two quantities $d_{\text{max}} = \min_{m=1,\ldots,M} d_{\text{max}}^{(m)}$ and $d_{+} = \min_{m=1,\ldots,M} d_{+}^{(m)}$.

For comparison purposes we consider the hypothetical case in which a single designer utilizes all covariates to design the observational study. This case is referred as the ``all-data designer''. The all-data designer will utilize the lasso logistic regression for estimating the propensity scores, and then subclassify or match the subjects accordingly. We compare the all-data designs with the best from the $M$ designers by the balance metrics $d_{\text{max}}^{(\text{all-data})} = \max_{j = 1, \ldots, p} \left\{d_{j}\left(S^{(\text{all-data})}\right)\right\}$ and $d_{+}^{(\text{all-data})} = \sum_{j = 1, \ldots, p} \mathbb{I}\left\{d_{j}\left(S^{(\text{all-data})}\right)  > 0.2\right\}$, respectively.
    
\subsection{Setting One Results}
\label{sec:simulation_results_1}

\begin{figure}[ht]
    \centering
    \includegraphics[width=\textwidth]{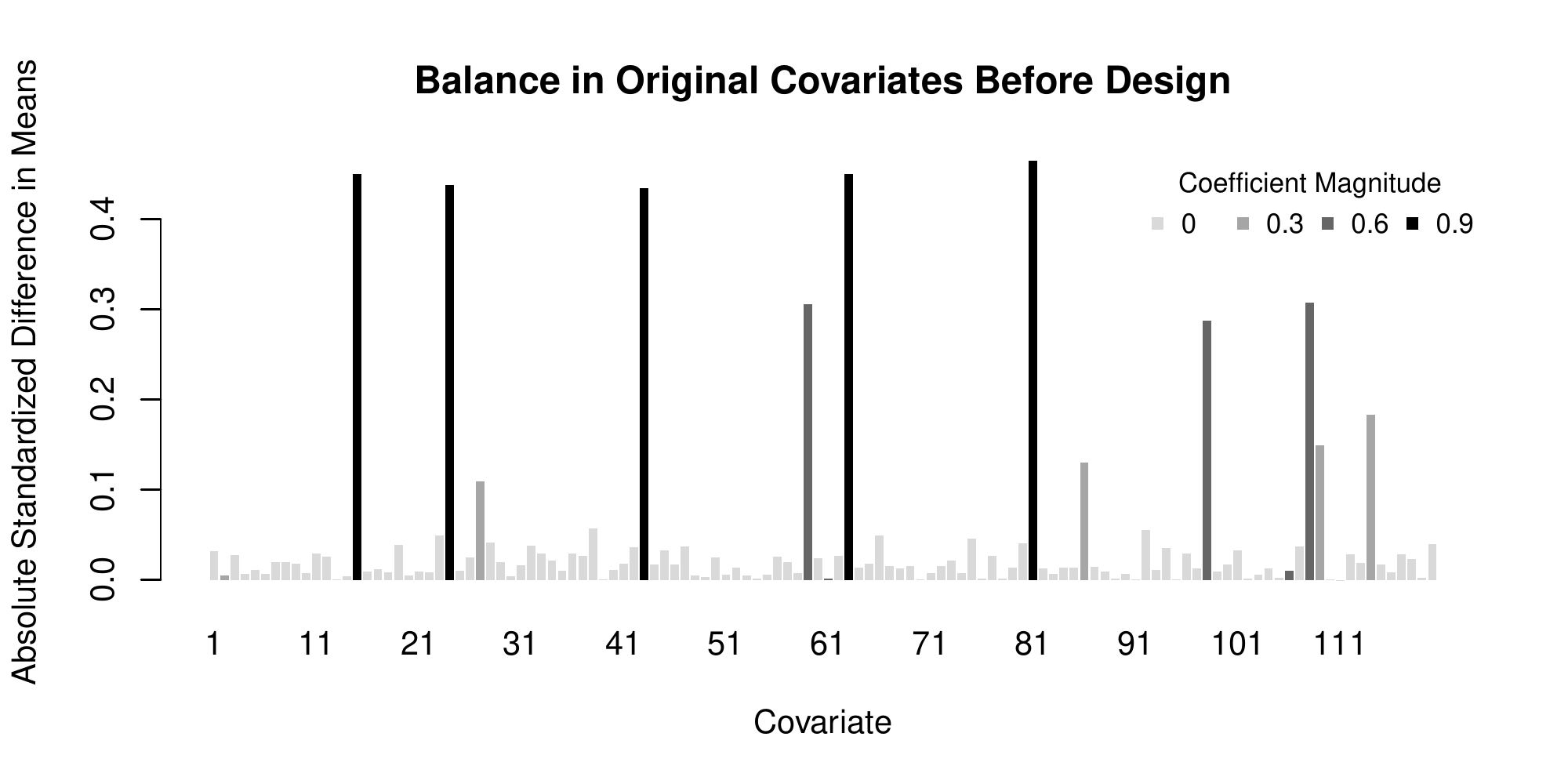}
    \caption{Covariate balance in one dataset, prior to design, simulated under the first setting. Each bar corresponds to one covariate. A coefficient value of $0$ indicates that the covariate was not selected in the treatment assignment mechanism.} 
    \label{fig:simu-initialBal-simu01}
\end{figure}

We first simulate 100 datasets under setting one. The range in the maximum absolute standardized differences in means between the treatment groups over all covariates across these datasets is $(0.35, 0.52)$. Figure \ref{fig:simu-initialBal-simu01} illustrates the covariate balance for one such dataset.

We observe that the designers' estimated propensity scores are highly correlated with the true propensity scores. Indeed, across the $100$ simulated datasets, the range of the correlations between the designers' estimated propensity scores and the true propensity scores is $(0.88, 0.99)$. Such high correlations and strong similarities between the designers' estimated propensity scores and the true values imply that the designers can recover substantial information on the true propensity scores by sharing conditional propensity scores with one another under our distributed design framework. 

\begin{figure}[ht]
    \centering
    \includegraphics[width=\textwidth]{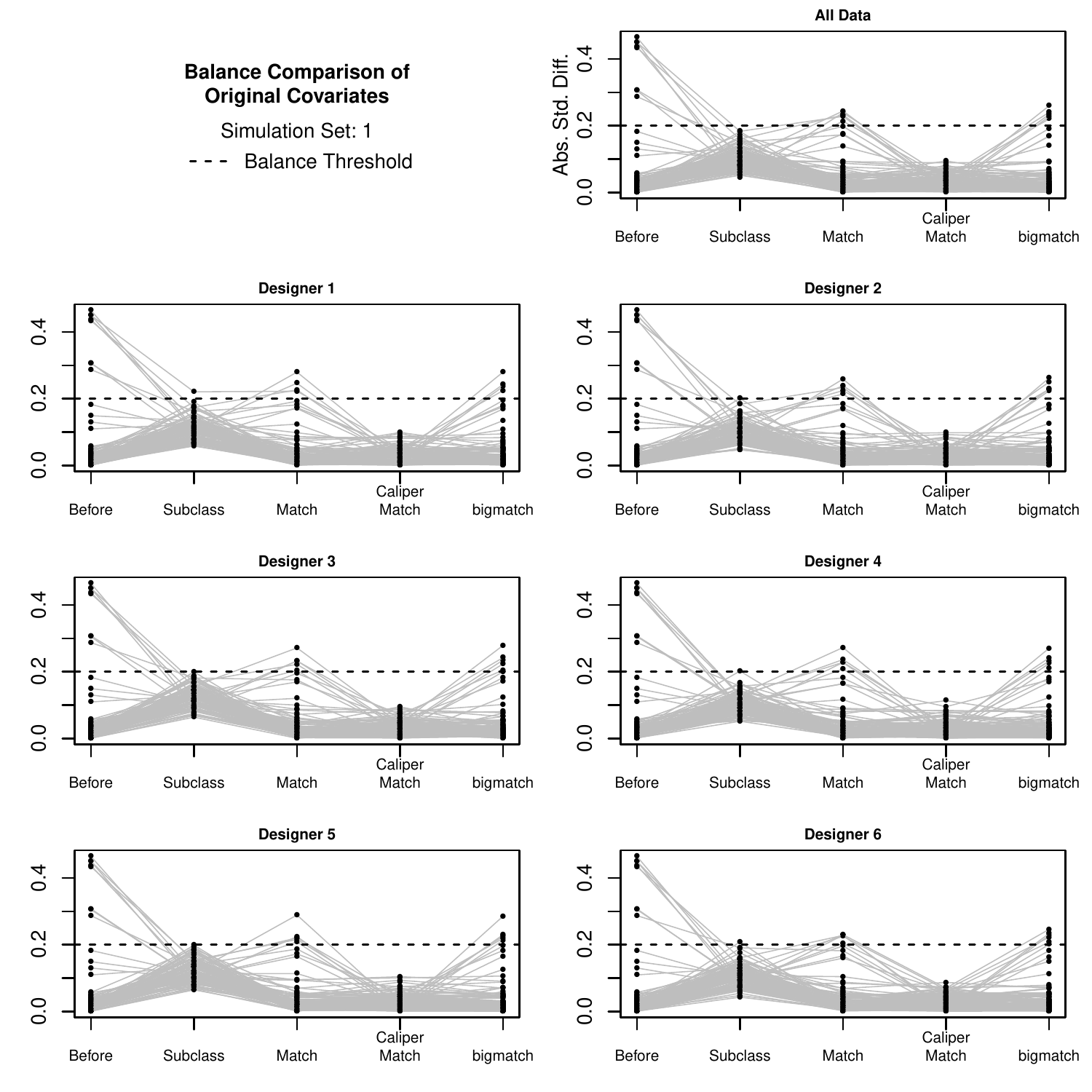}
    \caption{Comparison of the covariate balances for the simulated dataset considered in Figure \ref{fig:simu-initialBal-simu01} under the designs based on all the covariates, and under the designs by individual designers our distributed design framework. The four designs are subclassification (``subclass''), one-to-one nearest-neighbor matching, one-to-one matching with a caliper, and the optimal matching method by~\cite{YSR2020}.}
    \label{fig:simu-result-bal-01-main}
\end{figure}

Figure \ref{fig:simu-result-bal-01-main} illustrates how the designers improved covariate balance in different designs for the dataset considered in Figure \ref{fig:simu-initialBal-simu01}. The top-right panel shows the designs based on the estimated propensity scores using all the data, and the other panels correspond to the designers' proposals. Each line in a plot represents how the balance of one covariate changes, from the undesigned dataset to the different designs. Hence, a plot shows not only how the covariate balance are improved in general, but also how the different design methods are compared to each other under the default settings. 

We observe that the designers are able to improve the initially imbalanced covariates so that the almost all their after-design balance measures fall below the acceptance threshold of 0.2 via the subclassification and caliper matching designs. Comparing the designer-specific plots with the all-data plot, we see that the designers' improvements are similiar to the improvement by the all-data designer. The after-design balance measures of this dataset are presented in Table~\ref{tab:simu-result-bal-01-main}. We also notice that some covariates may have slightly worse after-design balance measures than before the design (represented by a line segment going upward from the "Before" column). They indicate the situation where improvement of poorly balanced covariates may require the introduction of acceptably minor imbalances for some other covariates. Such a phenomenon was also discussed by \citet[p.~12]{S2010}. 

The limited balance improvement via the matching designs (labeled by ``Match'' and ``bigmatch'') could be because some treated subjects are matched to control subjects that are not the closest, as our implementation of matching designs did not allow replacement of the control subjects. Nevertheless, the improvements of covariate balance achieved by the designers' proposals and the all-data approach via the matching designs are similar, implying that the distributed design framework could substitute the all-data approach in practice. The balance could be further improved by, for example, imposing a caliper on the propensity scores at the cost of discarding some subjects without close enough matching subjects.

\begin{figure}[ht]
    \centering
    \includegraphics[width=\textwidth]{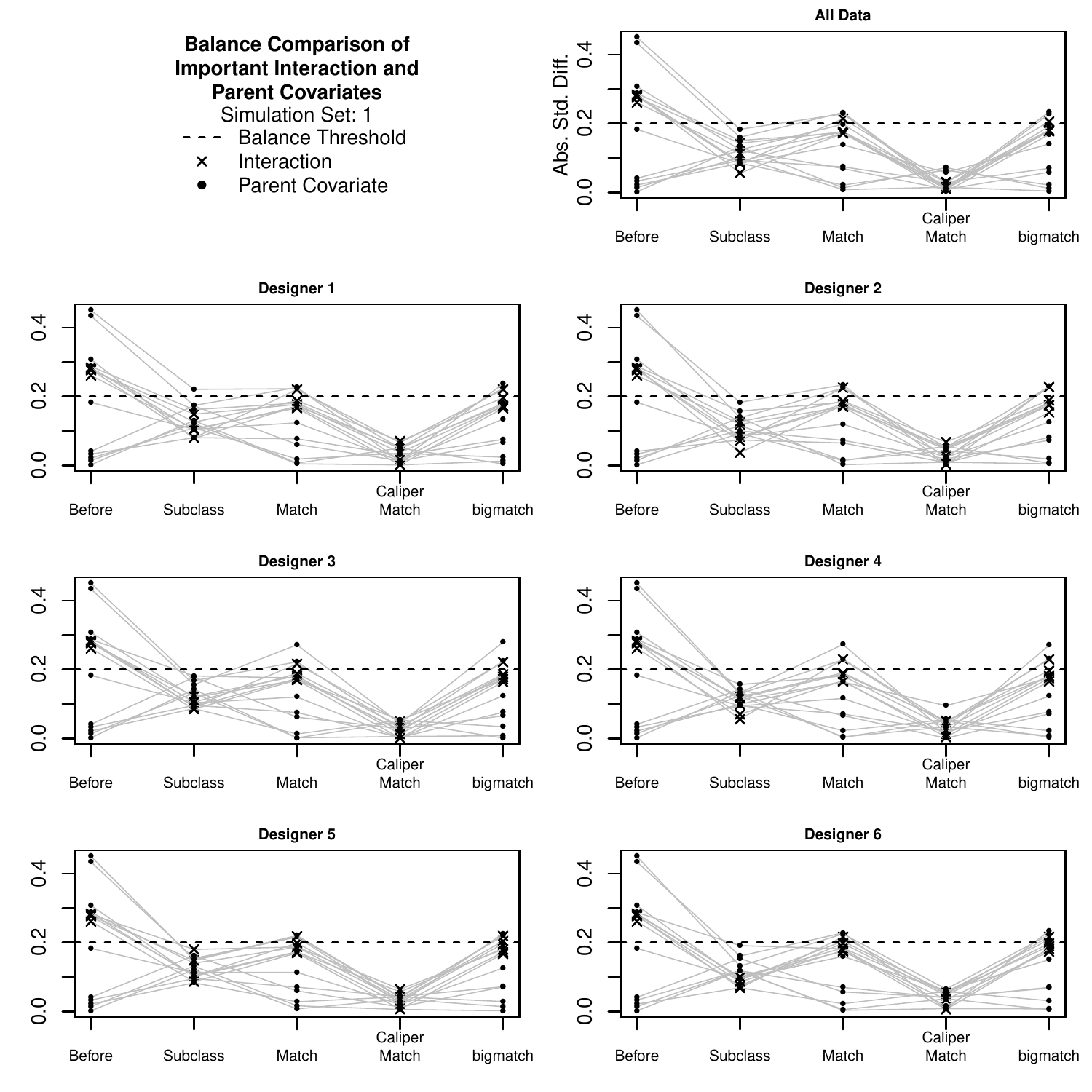}
    \caption{Comparison of balance values of the significant interactions and their parent covariates under the first simulation setting.}
    \label{fig:simu-result-bal-01-inter}
\end{figure}

To accompany Figure \ref{fig:simu-result-bal-01-main}, Figure \ref{fig:simu-result-bal-01-inter} demonstrates how the balance of the significant interactions as well as their parent covariates changed after design for the same simulated dataset. We see that all interactions have their balance levels below $0.2$ under the subclassification or caliper matching designs.

\begin{table}[ht]
\centering
\begin{tabular}{cc|rrrrrrr}
  \hline 
  \multirow{2}{*}{Metric} & \multirow{2}{*}{Design} & \multicolumn{7}{c}{Designer} \\
  & & All Data & M1 & M2 & M3 & M4 & M5 & M6 \\
  \hline 
  \multirow{4}{*}{\shortstack[c]{$d_{\text{max}}$ \\ (Initial: $.465$)}} & 
    Subclassification & .183 & .221 & .201 & .199 & .202 & \textbf{.198} & .208 \\
  & Matching          & .244 & .280 & .259 & .271 & .272 & .289 & \textbf{.230} \\
  & Caliper Matching  & .094 & .099 & .098 & .095 & .113 & .104 & \textbf{.087} \\
  & bigmatch          & .259 & .280 & .262 & .279 & .270 & .285 & \textbf{.246} \\
  \hline
  \multirow{4}{*}{\shortstack[c]{$d_{+}$ \\ (Initial: $8$)}} & 
    Subclassification & 0 & 1 & 1 & \textbf{0} & 1 & \textbf{0} & 1 \\
  & Matching          & 5 & \textbf{4} & 5 & 5 & 5 & 5 & 5 \\
  & Caliper Matching  & 0 & 0 & 0 & 0 & 0 & 0 & 0 \\
  & bigmatch          & 5 & \textbf{4} & 5 & 6 & 5 & 5 & 6 \\
  \hline
\end{tabular}
\caption{Covariate balance for the simulation dataset considered in Figure \ref{fig:simu-result-bal-01-main} from the first setting. The balance metrics are calculated based on the 120 covariates. The best after-design balance metrics achieved by the designers are marked in bold fonts.}
\label{tab:simu-result-bal-01-main}
\end{table}

\begin{figure}
    \centering
    \includegraphics[width=\textwidth]{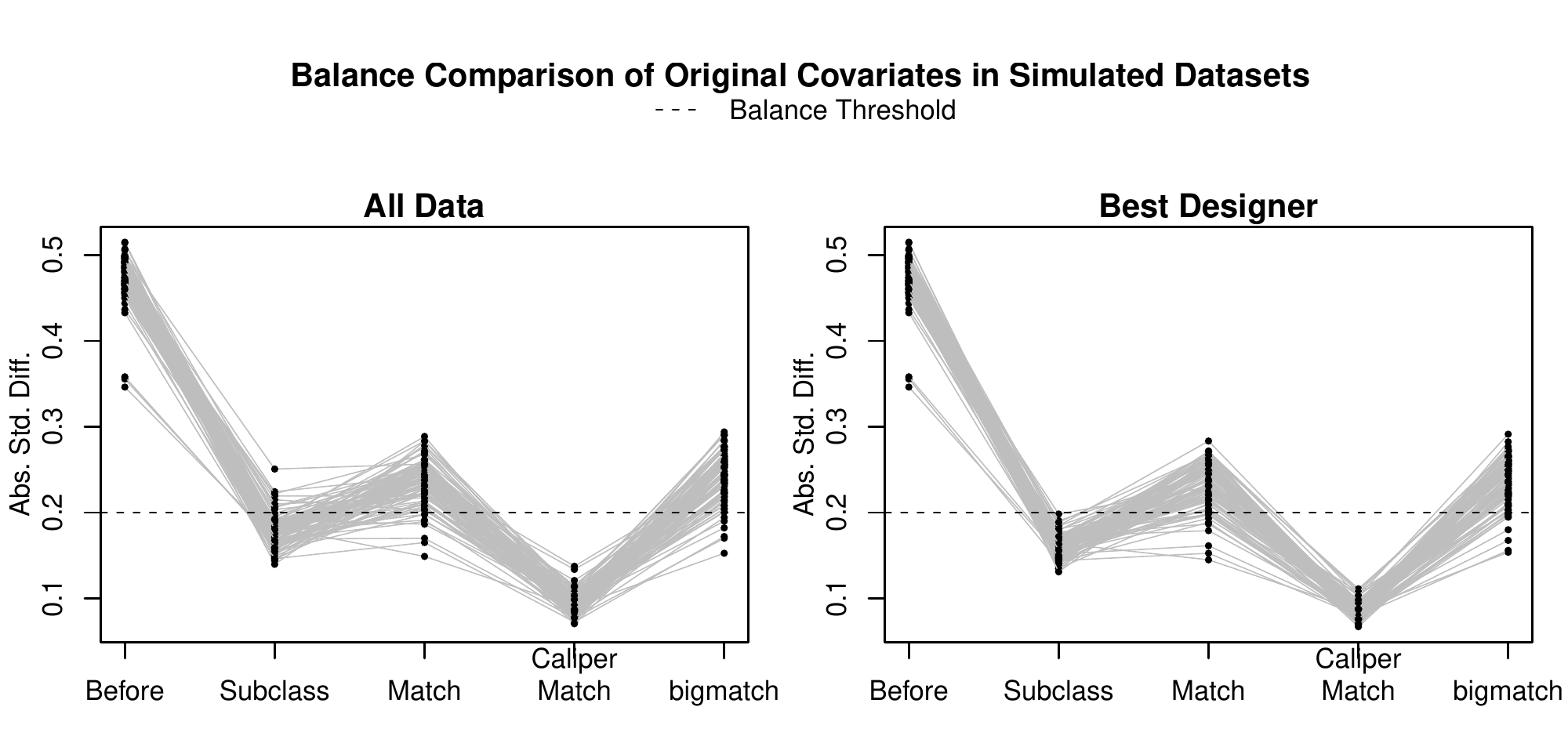}
    \caption{Change of balance in all simulated datasets from the first setting.}
    \label{fig:simu-result-bal-all-main}
\end{figure}

The patterns from Figure \ref{fig:simu-result-bal-01-main} are also present in the other $99$ simulated datasets. Figure~\ref{fig:simu-result-bal-all-main} compares the balance measure achieved by the all-data designer and the best designer via different designs. The comparable patterns between the two panels imply that the best designer under the distributed design framework manages to achieve the covariate balance similar to the all-data approach. Moreover, for the subclassification and the caliper matching designs, the best designer even slightly outperforms the all-data designer.

\subsection{Setting Two Results}
\label{sec:simulation_results_2}

\begin{figure}[ht]
    \centering
    \includegraphics[width=\textwidth]{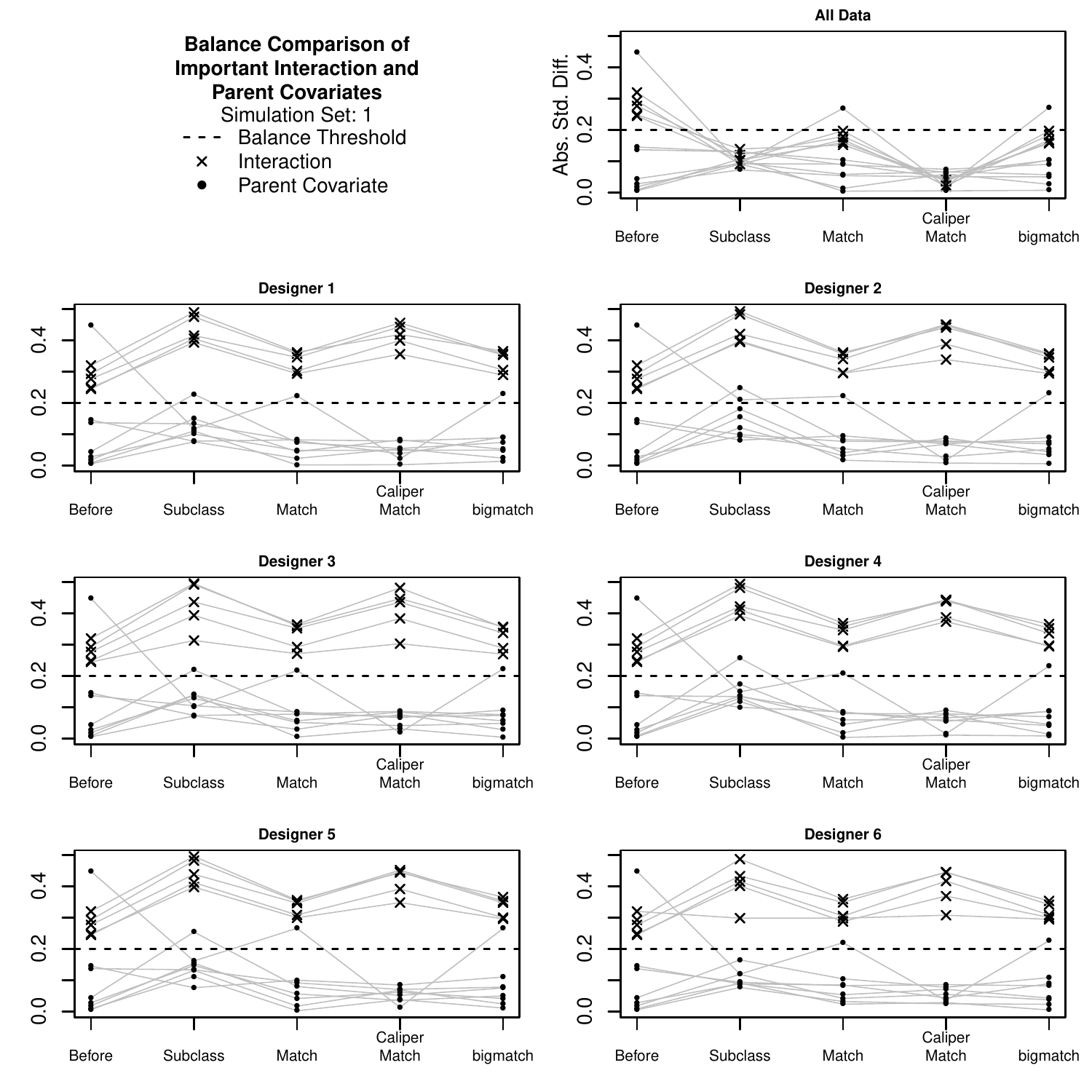}
    \caption{ Comparison of balance values of the significant interactions and their parent covariates for one dataset simulated under the second setting.}
    \label{fig:simu-result-bal-s2-01-inter}
\end{figure}

We next simulate $100$ datasets under setting two. We observe a similar pattern as in the first simulation setting, with the designers again improving the balance of the individual covariates to the same extent of the all-data designer. In contrast to the previous setting, we see that most of interaction terms actually had worse balance after design. Figure~\ref{fig:simu-result-bal-s2-01-inter} illustrates the case for one of the datasets. The worse balance of interaction terms is because no single matcher had the requisite granular information for achieving balance with respect to the interactions, although each designer had access to a subset of the covariates and the shared conditional propensity scores. Specifically, a parent covariate for a significant interaction could be contained in the conditional propensity score, but a designer would need the original values of that parent term to recover the interaction and achieve balance on it. Potential extensions for addressing this limitation are discussed in Section \ref{sec-conclusion}. 

\section{Demonstrations Based on the 2016 Atlantic Causal Inference Conference Data Challenge}
\label{sec-data}

We finally demonstrate the utility of our distributed design framework to reduce biases in causal inferences from Big Observational Data by utilizing the datasets from the 2016 ACIC Data Challenge \citep{aciccomp2016, DHSSC2019}. The contest organizers compiled multiple datasets consisting of $4802$ subjects and $58$ total covariates. The covariates were comprised of three categorical variables, five binary variables, $27$ count variables, and $23$ continuous variables. These covariates were developed from a real-life database. For each dataset compiled by the contest organizers, the subjects' treatment assignments and potential outcomes were simulated according to one of $77$ data generating processes. There were $100$ datasets simulated for each such setting. The $7700$ total datasets capture a wide variety of real-life scenarios for the design and analysis of Big Observational Data.

In this demonstration, we consider $M = 4$ designers, with $p^{(1)} = 15, p^{(2)} = 15, p^{(3)} = 15$, and $p^{(4)} = 13$. In addition, we only consider the $40$ of the $77$ total data generating processes that have more control subjects than treated subjects. The same set of design methods as in our previous simulation study was considered. To evaluate the performance of our distributed design framework in this demonstration, we again consider the ``all-data'' case of a single designer who has access to all covariates. Additionally, we consider the case of an ``oracle'' designer that directly subclassifies or matches the subjects using their true propensity scores, which were made available after the data challenge, so as to further assess the performance of our framework. As in the evaluations of our simulation studies, we use the absolute standardized difference in means between treatment groups as the primary covariate balance measure to evaluate the candidate designs. For the case of categorical covariates, we create indicator variables for the different levels and use the standardized difference for binary variables as the balance measure \citep{HSA2010}.

Our distributed design framework reduces the maximum absolute standardized difference in means between treatment groups to be less than or equal to $0.2$ for more than $3700$ of the $4000$ total datasets via the four types of designs implemented. Specifically, we group the datasets based on two of the factors involved in the data generating processes: covariate overlap condition and the type of treatment assignment model. A design of a dataset is regarded ``undesirable'' if the maximum absolute standardized difference of all the covariates is above the threshold $0.2$. For each group of datasets, we count the number of undesirable designs from different design methods and present the results in Table~\ref{tab:data-result-still-poor}. We observe that the best designer from the distributed design framework has the least number of undesirable designs compared to the hypothetical designers using the true propensity scores (``True PS'') or all the covariates (``All Data'') for all four design methods. The only exception occurs to the subclassification designs of datasets where there is a lack of covariate overlap between the treatment groups. The distributed design framework dominates the case where all the covariates are used simultaneously, while not reaching the balance level of subclassification designs that are based on the true propensity score. The reason is related to how the true propensity scores were generated. When there is a lack of overlap during the data generating process, subjects in a subspace of the covariates would have a large negative penalty to their linearized propensity scores and be forced into the control group. Due to their true propensity scores being extreme, these control subjects would be discarded in the subclassification design process by the ``oracle'' designer. 

\begin{table}[ht]
    \centering 
    \begin{tabular}{ccccrrr}
      \hline
      \begin{tabular}{@{}c@{}}Lack of \\ Overlap\end{tabular} & Trt Model & \begin{tabular}{@{}c@{}}Number of \\ Datasets\end{tabular} & Design & True PS & All Data & \begin{tabular}{@{}c@{}}Best \\ Designer\end{tabular} \\ 
      \hline
      \multirow{4}{*}{No}  & \multirow{4}{*}{Polynomial} &  \multirow{4}{*}{800} 
          & Subclass & 0 (.000) &  35 (.044) &  0 (.000) \\
      & & & Match &  5 (.006) &  2 (.002) &  2 (.002) \\
      & & & Caliper & 0 (.000) &  2 (.002) & 0 (.000) \\
      & & & bigmatch & 32 (.040) & 25 (.031) &  6 (.007) \\
      \hline 
      \multirow{4}{*}{No}  & \multirow{4}{*}{Step}       &  \multirow{4}{*}{800} 
          & Subclass & 1 (.001) &  59 (.074) &  0 (.000) \\ 
      & & & Match & 36 (.045) &  8 (.010) &  8 (.010) \\ 
      & & & Caliper & 1 (.001) &  0 (.000) & 0 (.000) \\ 
      & & & bigmatch & 50 (.062) & 42 (.052) & 10 (.013) \\ 
      \hline 
      \multirow{4}{*}{Yes} & \multirow{4}{*}{Polynomial} & \multirow{4}{*}{1200} 
          & Subclass & 4 (.003) & 238 (.198) & 95 (.079) \\ 
      & & & Match & 11 (.009) & 32 (.027) &  4 (.003) \\ 
      & & & Caliper & 10 (.008) & 47 (.039) & 1 (.001) \\ 
      & & & bigmatch & 66 (.055) & 101 (.084) & 28 (.023) \\ 
      \hline 
      \multirow{4}{*}{Yes} & \multirow{4}{*}{Step}       &  \multirow{4}{*}{800} 
          & Subclass & 3 (.004) & 165 (.206) & 65 (.081) \\ 
      & & & Match & 24 (.030) & 30 (.038) &  2 (.002) \\ 
      & & & Caliper & 21 (.026) & 32 (.040) & 2 (.002) \\ 
      & & & bigmatch & 65 (.081) & 69 (.086) & 14 (.018) \\ 
      \hline 
      \multirow{4}{*}{Yes} & \multirow{4}{*}{Linear}     &  \multirow{4}{*}{400} 
          & Subclass & 1 (.003) &  94 (.235) & 47 (.118) \\ 
      & & & Match & 11 (.028) & 13 (.032) &  0 (.000) \\ 
      & & & Caliper & 11 (.028) & 17 (.043) & 0 (.000) \\ 
      & & & bigmatch & 18 (.045) & 45 (.112) & 15 (.038) \\ 
      \hline
    \end{tabular}
    \caption{Counts of ``undesirable'' designs where the maximum absolute standardized difference is above $0.2$. The $4000$ ACIC datasets are grouped based on factors of the data generating process. For each of the four design methods, we obtain three designs using the true propensity score (``True PS''), the estimated propensity score based on all covariates simultaneously (``All Data''), and the distributed design framework.}
    \label{tab:data-result-still-poor}
\end{table}

\begin{figure}[ht]
    \centering
    \includegraphics[width=\textwidth]{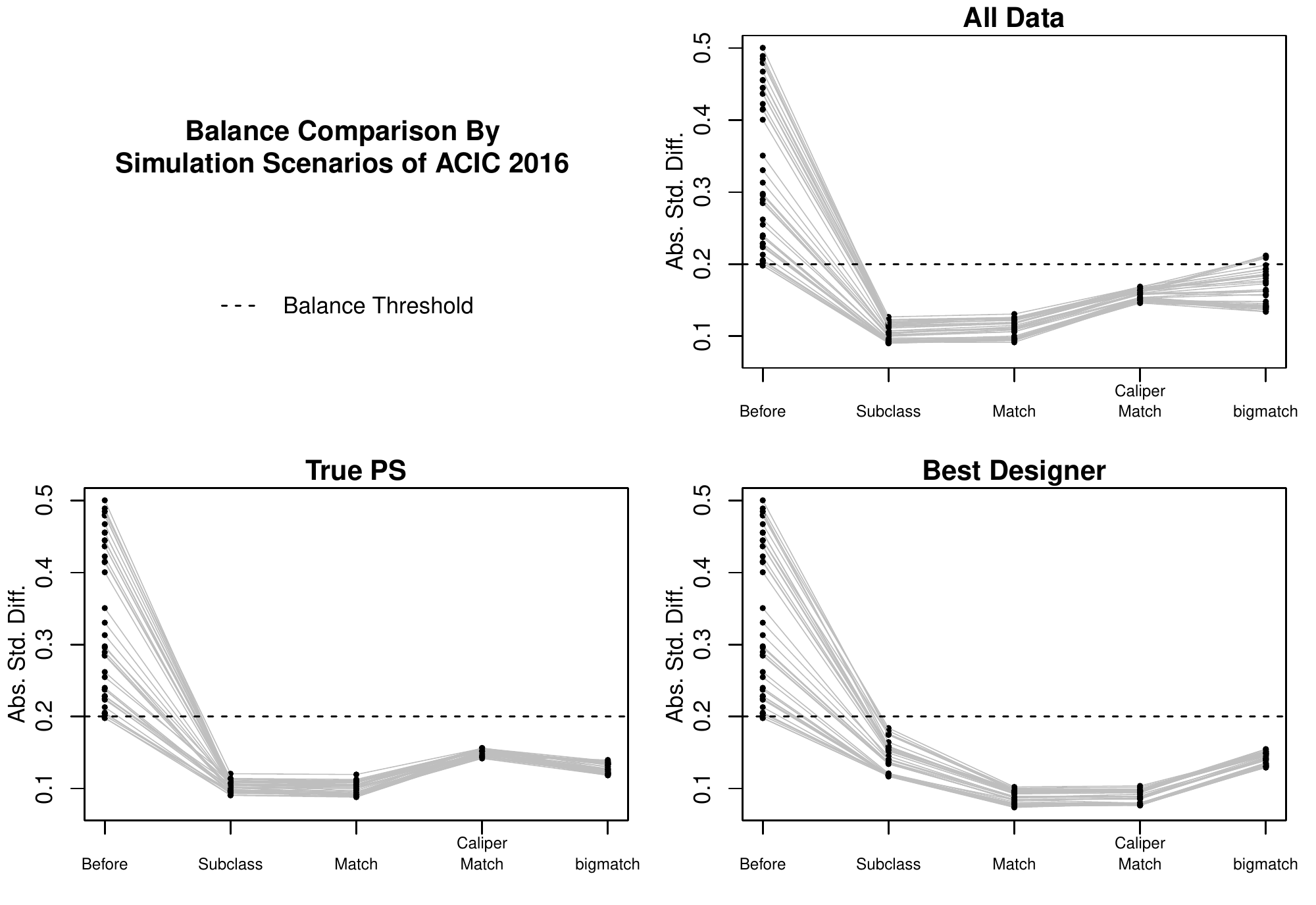}% 
    \caption{Covariate balance of datasets from the 2016 ACIC Data Challenge. Each line represents the average balance measures of datasets from one data generating process under different designs. The three panels show the results from the oracle designer who uses the true propensity score, the all-data designer who uses all the covariates, and the best designer from the distributed design framework.}
    \label{fig:data-result-bal-by-para}
\end{figure}

Figure~\ref{fig:data-result-bal-by-para} summarizes the improvement in covariate balance for datasets from different data generating processes. We compare the ``oracle'' designer who has access to the true propensity scores, the ``all-data'' designer utilizing all the covariates, and the best designer from our framework. For each of the $40$ data generating processes, represented by a line in the figure, the balance measures of the corresponding $100$ datasets are averaged and plotted. Again, the distributed design framework dominates the other two scenarios via matching designs, and achieves acceptable after-design balance via subclassification.

Both Table~\ref{tab:data-result-still-poor} and Figure~\ref{fig:data-result-bal-by-para} indicate that the distributed design framework can have better frequentist performance for achieving covariate balance with less computational effort and cognitive burden for the design process. We also note that the undesirable designs from our distributed framework can be avoided in real-life: The designers could more carefully examine and inform each other the overlap between treatment groups, or they could utilize propensity score models or design approaches that are more sophisticated or more targeted at the data and contexts.

\section{Concluding Remarks}
\label{sec-conclusion}

In this article we proposed a new framework for the distributed design of Big Observational Data. Our framework directly addresses the challenges of high dimensional and heterogeneous covariates, and the large volume of subjects, in Big Observational Data by dividing the subjects' covariates into multiple sets and assigning the sets of covariates to multiple designers. A key aspect of our framework is the sharing of low-dimensional summaries of the covariates across the designers. As demonstrated via multiple simulation studies and datasets from the 2016 ACIC Data Challenge, our framework can yield comparable, if not better, designs and covariate balance for observational studies compared to approaches in which a single designer utilizes all covariates to design the study. 

The four steps in our framework reduce the workload of each individual designer and overcome privacy and potential computational constraints in the design of a Big Observational Study. A great deal of flexibility exists for each step. The first step of assigning subsets of covariates to the designers could be guided under various principles. For example, the designers can work with a specific partition of the covariates that may be required by data security considerations or storage arrangements. Alternatively, for cases in which there are no privacy constraints, the covariates can be assigned to the designers based on the types of the covariates and the expertise levels of the designers. 
This first step is fundamental for the quality of the final designed study, both in terms of covariate balance and later causal inferences. The ideal covariate assignment should be based on identifying low-dimensional covariate spaces that capture the significant characteristics of the treated and control subjects, and assigning the pairs of covariates whose interactions are significant (for either the underlying treatment assignment mechanism or for explaining the variation in the potential outcomes) to individual designers who can then account for their interactions. 
The second step grants a great deal of freedom to the designers for estimating their conditional propensity scores and sharing their estimates with the other designers. For example, the conditional propensity scores can be estimated nonparametrically \citep{MRM2004}, or a designer can instead share their particular matching or subclassification of the subjects for the other designers' reference. Furthermore, no designer is necessarily restricted to share univariate summaries of the subjects' covariates with the other designers. Indeed, if the covariates are thought to be better summarized in a two-dimensional space (e.g., via a principal component analysis), then in the second step of our framework a designer can in fact share the two-dimensional covariate summaries with the other designers. More generally, the designers can summarize their covariates into a lower-dimensional balancing score \citep{RR1983} that is then shared with the other designers. 
In the third step, each designer may choose a design method that would achieves the best balance possible. In addition to the subclassification and matching designs that we implemented, other options include, for example, variable ratio matching~\citep{MR2001}, full matching~\citep{R1991}, or optimization-motivated methods~\citep{Z2012,BVZ2020}.

Additional pre-processing steps can be added to our framework to improve its efficacy in designing Big Observational Data. For the first step, a computer-based pilot inspection of all covariates for a random subset of the subjects can be implemented before assigning the covariates to the designers. This can assess several important, pre-specified balance measures for the covariates along with their transformations and interactions. The information can then facilitate the division of the covariates and implementation of the other steps in the framework by identifying those covariates that are likely to have the worst balance or insufficient overlap across the treatment groups for the entire dataset. Other pilot studies can be performed under our framework to yield preliminary knowledge about the covariates, and the designers can use the combination of their domain expertise with such preliminary information on the data to better design the observational study.

When evaluating candidate designs, a variety of balance measures can be considered beyond the common standardized difference in means between treatment groups. Assessing covariate balance in practice is a step that requires a great deal of judgement and wisdom. Furthermore, the designers would have to prioritize the balancing of different covariates. An advantage of our framework is that by assigning only a subset of covariates to a designer, each designer can better use their domain expertise, computational and cognitive capacity, and prudence to check the balance of their assigned covariates without being overwhelmed by the extremely large number of covariates common in modern Big Observational Data. 

Several aspects of the distributed design framework remain to be investigated. One feature is how the correlations among covariates should guide the assignment of covariates to the different designers in the first step of the framework, and how the correlations affect the quality of the candidate designs with respect to covariate balance. Another is enhancing the framework by having each designer efficiently summarize their assigned covariates into a lower-dimensional vector besides the conditional propensity score that accurately represents the subjects' covariates, and can by used by the other designers to obtain improved designs. An interesting question for this enhancement is how a balancing vector can be defined based on the information shared across multiple designers. A final aspect is exploring other approaches in which the designers can collaborate during the design process. One possibility is to have each designer examine those subjects that are not well-matched by the other designers, so as to discover patterns in the covariates that may not have been obvious at the start of the design process. Designers can conduct multiple rounds of adjusting and discussing their candidate designs with one another before reaching the final designed study.

% Acknowledgements should go at the end, before appendices and references

\acks{We would like to acknowledge the support for this project from Purdue University ITaP Explanatory Modeling Project Grant.}

\appendix

\section{Details of Design Methods}\label{appen:design-method}

The particular propensity score-based design methods that we will consider in our simulation studies and demonstrations include one-to-one nearest neighbor matching, nearest neighbor matching with caliper, and the technique recently developed by \citet{YSR2020} to conduct optimal matching for datasets with large numbers of experimental units. The nearest neighbor matching will for each treatment unit find one control unit that has the closest propensity score. The version of matching with caliper will search for the matching unit only from the control units whose propensity scores are no further than the caliper range from the treatment unit's score. In order to reduce the computational complexity of finding matched pairs from all the available units, \citet{YSR2020} first impose an optimally chosen caliper of the propensity score to eliminate ineligible poor matched pairs, and then identify the optimal matching from the reduced, tractable solution space. 

Besides matching methods, we also implement the iterative subclassification procedure based on estimated propensity scores developed by \citet[p.~290--294]{IR2015}. This procedure commences with the entire dataset as the initial subclass. In each iteration of this procedure, the researcher considers a hypothetical split at the median propensity score of an existing subclass. The subclass will be divided into two subclasses if (1) the $t$-test comparing the mean propensity scores of the treated and control subjects in the current subclass gives a $p$-value below a pre-specified level (e.g., 0.15), (2) either one of the hypothetical subclasses exceeds the pre-specified subclass size, and (3) in either of the hypothetical subclasses, the size of each treatment group exceeds a pre-specified group size. This procedure ends when no subclass can be further split. In general, this procedure can accommodate a flexible number of subclasses. The last two conditions ensure that the final subclasses are sufficiently large for statistical analyses.

\section{Data Generating Mechanism for Simulation Study}\label{appen:simu-data}

Below we provide the details of the data generating mechanism for our simulation study.
First, $15$ covariates are randomly selected to be active in the specification. Second, each selected covariate is raised to the second or third power (and further standardized) with probability $0.5$ to add nonlinear covariate effects in the treatment assignment. Third, the selected covariates are randomly assigned into three equal-sized groups, with each group corresponding to different magnitudes of coefficients for the covariates in the specification of $\mathrm{log}\{p_i/(1-p_i)\}$. The specific coefficients for the covariates in the three groups are $\pm0.3$, $\pm0.6$, and $\pm0.9$, respectively, with the signs of the coefficients being selected at random. Fourth, five two-way covariate interaction terms are randomly created for the specification. Each interaction term will have at least one parent covariate belonging to the selected main effects, corresponding to the weak effect heredity principle \citep[p.~173]{WH2009}. The coefficients for the interaction terms are $\pm 0.6$, with the sign selected at random. Fifth, the intercept term of the specification is set to $\log\{0.2/(1-0.2)\} \approx -1.386$. The aforementioned terms in the five steps are linearly combined based on their coefficients for $\log\{p_{i}/(1 - p_{i})\}$ and thus the probability of treatment assignment $p_{i}$.

The random selection and transformation of covariates in the first step results in each designer having different types of information when creating their proposal. The selected low, medium, and high coefficient magnitudes in the third step will introduce small to large imbalances in the selected covariates for the simulated datasets. Our choice of the intercept term in the fifth step implies that for a subject whose (standardized) covariate values are all zero (i.e., their covariate values correspond to the average covariate values among all the subjects), their probability of being assigned the active treatment is approximately $0.2$. In addition, it implies that approximately $20\%$ of the subjects in each dataset will be in the treatment group. This is consistent with the usual situation in real-life studies in which the treatment group is a minority in the dataset. 

Equation (\ref{simu-data-PS-eg}) provides an example of a specification for $\mathrm{log}\{p_i/(1-p_i)\}$. In this equation $g(\cdot)$ indicates standardization of a term inside the parentheses. This specification corresponds to the first simulation setting in which both parent covariates of an interaction are assigned as a pair to the same designer. We write out the terms in the specification according to the designers' covariate assignments. For example, covariates assigned to designer $1$ include those in the first row (namely, $9$, $11$, $13$, and $15$). Designers $5$ and $6$ have covariates whose coefficients are large in magnitude (namely, $0.6$ and $0.9$, respectively).
\begin{equation}
\label{simu-data-PS-eg}
\begin{aligned}
    \log&\left(\frac{p_i}{1 - p_i}\right)
    = -1.4 \\
    & - 0.3 X_{i,13} - 0.6 g\left(X_{i,15}^{3}\right) - 0.6 X_{i,9}X_{i,15} - 0.6 X_{i,11}X_{i,13} \\
    & - 0.3 X_{i,22} + 0.9 X_{i,25} - 0.9 g\left(X_{i,26}^{3}\right) - 0.6 g\left(X_{i,35}^{3}\right) - 0.6 X_{i,21}X_{i,25} \\
    & + 0.6 g\left(X_{i,42}^{2}\right) + 0.3 X_{i,58} + 0.6 X_{i,50}X_{i,58} \\
    & - 0.3 g\left(X_{i,64}^{3}\right) - 0.3 X_{i,71} + 0.6 X_{i,62}X_{i,64} \\
    & + 0.6 X_{i,85} - 0.9 X_{i,86} \\
    & - 0.6 X_{i,106} + 0.9 X_{i,110} - 0.9 X_{i,118}
\end{aligned}
\end{equation}

\vskip 0.2in
\bibliography{mybib}

\end{document}